\PassOptionsToPackage{unicode}{hyperref}
\PassOptionsToPackage{hyphens}{url}
\PassOptionsToPackage{dvipsnames,svgnames,x11names}{xcolor}
\documentclass[
  12pt]{article}
\usepackage{amsfonts}
\newcommand{\cN}{\mathcal{N}}
\newcommand{\cA}{\mathcal{A}}
\newcommand{\cC}{\mathcal{C}}
\newcommand{\cS}{\mathcal{S}}
\newcommand{\cF}{\mathcal{F}}
\usepackage{amsmath,amssymb}
\usepackage{amsthm}
\usepackage{iftex}
\ifPDFTeX
  \usepackage[T1]{fontenc}
  \usepackage[utf8]{inputenc}
\else
  \usepackage{fontspec}
\fi
\usepackage{lmodern}
\ifPDFTeX\else  
\fi
\IfFileExists{upquote.sty}{\usepackage{upquote}}{}
\IfFileExists{microtype.sty}{
  \usepackage[]{microtype}
  \UseMicrotypeSet[protrusion]{basicmath} 
}{}
\makeatletter
\@ifundefined{KOMAClassName}{
  \IfFileExists{parskip.sty}{%
    \usepackage{parskip}
  }{
    \setlength{\parindent}{0pt}
    \setlength{\parskip}{6pt plus 2pt minus 1pt}}
}{
  \KOMAoptions{parskip=half}}
\makeatother
\usepackage{xcolor}
\makeatletter
\ifx\paragraph\undefined\else
  \let\oldparagraph\paragraph
  \renewcommand{\paragraph}{
    \@ifstar
      \xxxParagraphStar
      \xxxParagraphNoStar
  }
  \newcommand{\xxxParagraphStar}[1]{\oldparagraph*{#1}\mbox{}}
  \newcommand{\xxxParagraphNoStar}[1]{\oldparagraph{#1}\mbox{}}
\fi
\ifx\subparagraph\undefined\else
  \let\oldsubparagraph\subparagraph
  \renewcommand{\subparagraph}{
    \@ifstar
      \xxxSubParagraphStar
      \xxxSubParagraphNoStar
  }
  \newcommand{\xxxSubParagraphStar}[1]{\oldsubparagraph*{#1}\mbox{}}
  \newcommand{\xxxSubParagraphNoStar}[1]{\oldsubparagraph{#1}\mbox{}}
\fi
\makeatother

\usepackage{longtable,booktabs,array}
\usepackage{calc} 
\usepackage{etoolbox}
\makeatletter
\patchcmd\longtable{\par}{\if@noskipsec\mbox{}\fi\par}{}{}
\makeatother
\IfFileExists{footnotehyper.sty}{\usepackage{footnotehyper}}{\usepackage{footnote}}
\makesavenoteenv{longtable}
\usepackage{graphicx}
\makeatletter
\def\maxwidth{\ifdim\Gin@nat@width>\linewidth\linewidth\else\Gin@nat@width\fi}
\def\maxheight{\ifdim\Gin@nat@height>\textheight\textheight\else\Gin@nat@height\fi}
\makeatother
\setkeys{Gin}{width=\maxwidth,height=\maxheight,keepaspectratio}
\makeatletter
\def\fps@figure{htbp}
\makeatother

\makeatletter
\@ifpackageloaded{caption}{}{\usepackage{caption}}
\AtBeginDocument{%
\ifdefined\contentsname
  \renewcommand*\contentsname{Table of contents}
\else
  \newcommand\contentsname{Table of contents}
\fi
\ifdefined\listfigurename
  \renewcommand*\listfigurename{List of Figures}
\else
  \newcommand\listfigurename{List of Figures}
\fi
\ifdefined\listtablename
  \renewcommand*\listtablename{List of Tables}
\else
  \newcommand\listtablename{List of Tables}
\fi
\ifdefined\figurename
  \renewcommand*\figurename{Figure}
\else
  \newcommand\figurename{Figure}
\fi
\ifdefined\tablename
  \renewcommand*\tablename{Table}
\else
  \newcommand\tablename{Table}
\fi
}
\@ifpackageloaded{float}{}{\usepackage{float}}
\floatstyle{ruled}
\@ifundefined{c@chapter}{\newfloat{codelisting}{h}{lop}}{\newfloat{codelisting}{h}{lop}[chapter]}
\floatname{codelisting}{Listing}

\makeatother
\makeatletter
\@ifpackageloaded{caption}{}{\usepackage{caption}}
\@ifpackageloaded{subcaption}{}{\usepackage{subcaption}}
\makeatother

\ifLuaTeX
  \usepackage{selnolig}  
\fi
\usepackage{natbib}
\usepackage{bookmark}

\IfFileExists{xurl.sty}{\usepackage{xurl}}{} 
\hypersetup{
  pdftitle={Randomization Inference for Local Projections with Persistent Instruments in Short Panels},
  pdfauthor={Pedro Picchetti},
  pdfkeywords={Local projections, Instrumental variables, Randomization inference, Panel data, Persistent instruments, Nuisance parameters},
  colorlinks=true,
  linkcolor={blue},
  filecolor={Maroon},
  citecolor={Blue},
  urlcolor={Blue},
  pdfcreator={LaTeX via pandoc}}

\usepackage{algorithm}
\usepackage{algpseudocode}

\newcommand{\Cov}{\mathbb{C}\mathrm{ov}}

\newcommand{\Prob}{\mathbb{P}}

\newcommand{\PR}{\mathbb{P}_{\mathcal R}}
\newcommand{\ER}{\mathbb{E}_{\mathcal R}}
\newcommand{\VR}{\mathbb{V}_{\mathcal R}}
\newcommand{\CovN}{\operatorname{Cov}_{N}}
\newcommand{\hatt}{\widehat{\tau}}
\newcommand{\RF}{\mathrm{RF}}
\newcommand{\FS}{\mathrm{FS}}
\newcommand{\AR}{\mathrm{AR}}
\newcommand{\one}{\mathbf{1}}
\newcommand{\dto}{\xrightarrow{d}}
\newcommand{\pto}{\xrightarrow{p}}

\newcommand{\anon}{1}
\newtheorem{proposition}{Proposition}
\newtheorem{corollary}{Corollary}
\newtheorem{lemma}{Lemma}
\newtheorem{theorem}{Theorem}
\newtheorem{assumption}{Assumption}
\theoremstyle{definition}

\theoremstyle{remark}
\newtheorem{remark}{Remark}

\begin{document}

\def\spacingset#1{\renewcommand{\baselinestretch}%
{#1}\small\normalsize} \spacingset{1}


\if1\anon
{
  \title{\bf Design-Based Inference for Fuzzy Difference-in-Differences}
  \author{Pedro Picchetti\thanks{
    pedro.picchetti@uc.cl.}\hspace{.2cm}\\
    Institute of Economics, PUC-Chile\\
    }
  \maketitle
} \fi

\if0\anon
{
  \bigskip
  \bigskip
  \bigskip
  \begin{center}
    {\LARGE\bf Design-Based Inference for Fuzzy Difference-in-Differences}
\end{center}
  \medskip
} \fi

\bigskip
\begin{abstract}
I study fuzzy difference-in-differences designs from a design-based perspective, in which potential outcomes and treatment decisions are fixed and only the assignment of a policy is random, with heterogeneous and possibly unknown assignment probabilities. The Wald-DiD design estimand equals an assignment-weighted complier effect plus contamination from switchers, from changes in always-takers' effects and from trends in untreated outcomes. I derive the conditions under which the estimand holds a clear causal interpretation. The conditions are weaker than parallel trends in both outcomes and treatments. We propose an Anderson--Rubin-type test with a conservative Neyman variance that remains valid under weak first stages, a sensitivity analysis for the complier effect, and estimators for designs with known assignment probabilities. An application to the Oregon Health Insurance Experiment illustrates the results. 
\end{abstract}

\noindent%
\vfill

\newpage
\spacingset{1.8} 

\section{Introduction}\label{sec:intro}

Difference-in-differences (DiD) is a popular method for causal inference among applied social scientists. In its canonical format, there are two time periods. In the first period, all units are untreated, and in the second period, some units take treatment and other remain untreated. Under the parallel trends assumption, the DiD estimand identifies the average treatment effect on the treated (ATT). However, many difference-in-differences applications are \emph{fuzzy}: a policy changes the incentive to take a treatment, but take-up is imperfect. Some units are treated before the policy and start treatment for reasons unrelated to it, while some units are exposed to the policy and remain untreated. The canonical estimator in this setting is the Wald-DiD ratio, which divides the DiD in outcomes by the DiD in treatment rates \citep{Duflo2001}. \citet{deChaisemartinDHaultfoeuille2018} showed that, in the conventional sampling-based framework, the Wald-DiD estimand identifies a local average treatment effect only under strong homogeneity restrictions: parallel trends must be combined with stability of treatment effects over time and homogeneity of effects between groups.

A growing literature argues that, in many DiD applications, uncertainty is more naturally attributed to the assignment of the policy than to sampling from a superpopulation \citep{AbadieAtheyImbensWooldridge2020,AtheyImbens2022,AbadieAtheyImbensWooldridge2023,RothSantAnna2023JPE}. \citet{RambachanRoth2026} develop a design-based framework for quasi-experiments in which units have heterogeneous and unknown assignment probabilities, so that selection into treatment is permitted, and show how the biases of popular estimators can be expressed as finite-population covariances between assignment probabilities and potential outcomes. This paper brings that perspective to the fuzzy DiD framework.

I consider a two-period panel in which a policy (the \emph{assignment}) becomes available in the second period. Treatment status is binary and monotone in both assignment and time, which partitions the population into never-takers, always-takers, compliers and switchers (units that adopt treatment in the second period irrespective of assignment). 

I derive an exact causal decomposition of the design-based Wald-DiD estimand . It equals an assignment-probability-weighted average of complier effects plus three contamination terms: the effects of treatment on switchers, the change over time in the effects of treatment on always-takers, and trends in untreated outcomes. Each term enters the expression through its finite-population covariance with the assignment probabilities. All three vanish under complete randomization, so in the design-based framework the problems highlighted by \citet{deChaisemartinDHaultfoeuille2018} arise  from the interaction between heterogeneity in assignment probabilities interacting and heterogeneity in effects and trends. 

Second, I characterize the conditions under which the Wald-DiD design estimand equals the weighted complier effect: the assignment probabilities must be uncorrelated with a single contamination index that combines untreated trends, switcher effects and the drift in always-takers' effects. This condition is necessary and sufficient, and it is strictly weaker than the pair of orthogonality conditions (design-based parallel trends for outcomes \emph{and} for treatment) that would follow from applying existing results directly \citep{hudson2017interpreting,RambachanRoth2026}. For example, it holds under design-based parallel trends and homogeneous effects even when switchers are concentrated among units likely to be assigned.

Third, I propose an Anderson--Rubin-type test  based on a difference in means of the null-adjusted outcome change, studentized by a Neyman-type variance estimator. The test is asymptotically valid under the rejective-sampling design of \citet{RambachanRoth2026}, with a conservative variance estimator whose slack we characterize, and its validity does not depend on the strength of the first stage. Inverting the test gives Fieller-type confidence sets, which are unbounded exactly when the first stage is not significant.

Fourth, I develop a sensitivity analysis for the complier effect. Bounds on the two contamination covariances deliver a closed-form identified set and a confidence set that is again robust to weak identification. Tests of the null of no complier effect depend only on the reduced-form contamination bound, which yields a simple breakdown value. When the assignment probabilities are known, as in stratified experiments and lotteries, the contamination terms are themselves estimable and the complier effect is point identified.

The results are illustrated with the Oregon Health Insurance Experiment \citep{Finkelstein2012,Taubman2014}, in which the right to apply for Medicaid was allocated by lottery and compliance was imperfect: some individuals who were not selected enrolled through other routes, and selection probabilities differed by household size. Because the probabilities are known, the application serves as a benchmark: it measures the contamination that an analyst ignoring the heterogeneity in assignment probabilities would face.

\paragraph{Related literature}
The paper contributes to the literature on DiD with heterogeneous effects (\citealp{deChaisemartinDHaultfoeuille2018,deChaisemartinDHaultfoeuille2020,CallawaySantAnna2021,GoodmanBacon2021,SunAbraham2021,BorusyakJaravelSpiess2024}; see \citealp{deChaisemartinDHaultfoeuille2023survey} and \citealp{RothSantAnnaBilinskiPoe2023} for reviews), and specifically to fuzzy DiD \citep{deChaisemartinDHaultfoeuille2018}. It also contributes to design-based inference \citep{Neyman1990,ImbensRubin2015,LiDing2017,AbadieAtheyImbensWooldridge2020,AtheyImbens2022,RothSantAnna2023JPE,RambachanRoth2026}, in particular for IV \citep{ImbensRosenbaum2005,KangPeckKeele2018}. We share the assignment model of \citet{RambachanRoth2026}, who analyse two-stage least squares under strong-instrument asymptotics. Our contributions relative to their work are the fuzzy DiD decomposition and characterization, inference that is robust to weak identification \citep{StaigerStock1997,AndrewsStockSun2019}, and a sensitivity analysis that bounds the first-stage and reduced-form contamination jointly, in the spirit of \citet{Rosenbaum2002}, \citet{ManskiPepper2018}, \citet{CinelliHazlett2020} and \citet{RambachanRoth2023}.

The rest of the paper is organized as follows: Section~\ref{sec:setup} introduces the framework. Section~\ref{sec:ident} presents the identification results, Section~\ref{sec:inference} inference, and Section~\ref{sec:sensitivity} the sensitivity analysis and the case of known assignment probabilities. Section~\ref{sec:sim} summarizes simulations and Section~\ref{sec:app} presents the application. Proofs of the main results are in Appendix~\ref{app:proofs}. The Online Appendix contains the simulation design, additional simulation results and the proof of an auxiliary lemma.

\section{Setup}\label{sec:setup}

\subsection{Potential outcomes and assignment}

We observe a balanced panel of $N$ units, $i=1,\dots,N$, in two periods, $t\in\{0,1\}$. A policy becomes available in period $1$. Let $Z_i\in\{0,1\}$ indicate whether unit $i$ is assigned to the policy in period $1$; no unit is assigned in period $0$. Treatment is binary. We write $D_{i0}$ for treatment status in period $0$ and $D_{i1}(z)$ for potential treatment status in period $1$ under assignment $z$. Potential outcomes are $Y_{it}(d)$, $d\in\{0,1\}$. This notation, which is closer to the IV literature than that of \citet{deChaisemartinDHaultfoeuille2018}, encodes two maintained restrictions.

\begin{assumption}[Exclusion and no anticipation]\label{ass:exclusion}
Assignment affects outcomes only through treatment in both periods, and period-$0$ treatment and outcomes do not depend on period-$1$ assignment, so that the observed data are $D_{i0}$, $Y_{i0}=Y_{i0}(D_{i0})$, $D_{i1}=D_{i1}(Z_i)$ and $Y_{i1}=Y_{i1}(D_{i1}(Z_{i}))$.
\end{assumption}

Let $\tau_{it}=Y_{it}(1)-Y_{it}(0)$ be the unit-level effect in period $t$, $\Delta\tau_i=\tau_{i1}-\tau_{i0}$ its change over time, and $\delta_i=Y_{i1}(0)-Y_{i0}(0)$ the untreated outcome trend. The first differences, as functions of assignment, are
\begin{equation}\label{eq:potential-differences}
\Delta D_i(z)=D_{i1}(z)-D_{i0},\qquad \Delta Y_i(D_{i}(z))=Y_{i1}\bigl(D_{i1}(z)\bigr)-Y_{i0}(D_{i0}),\qquad z\in\{0,1\},
\end{equation}
and the observed differences are $\Delta D_i=\Delta D_i(Z_i)$ and $\Delta Y_i=\Delta Y_i(D_{i}(Z_i))$.

\begin{assumption}[Monotonicity in assignment and time]\label{ass:monotone}
For all $i$, (a) $D_{i1}(1)\ge D_{i1}(0)$; (b) $D_{i1}(z)\ge D_{i0}$ for $z\in\{0,1\}$.
\end{assumption}

Part (a) is the usual IV monotonicity condition \citep{ImbensAngrist1994}. Part (b) rules out treatment discontinuation. Together they leave four treatment paths, which partition the population into the groups in Table~\ref{tab:types}. In particular, always-takers are exactly the units with $D_{i0}=1$, and their number $N_{\mathcal{A}}=\sum_i D_{i0}$ is observed.

\begin{table}[t]
\caption{\label{tab:types}Compliance types under Assumption~\ref{ass:monotone}.}
\begin{center}
\begin{tabular*}{\textwidth}{@{\extracolsep{\fill}}lcccccc@{}}
\hline\hline
Type & $D_{i0}$ & $D_{i1}(0)$ & $D_{i1}(1)$ & $\Delta D_i(0)$ & $\Delta Y_i(D_{i}(0))$ & $\Delta Y_i(D_{i}(1))-\Delta Y_i(D_{i}(0))$\\
\hline
Never-takers $\cN$ & 0 & 0 & 0 & 0 & $\delta_i$ & 0\\
Always-takers $\cA$ & 1 & 1 & 1 & 0 & $\delta_i+\Delta\tau_i$ & 0\\
Compliers $\cC$ & 0 & 0 & 1 & 0 & $\delta_i$ & $\tau_{i1}$\\
Switchers $\cS$ & 0 & 1 & 1 & 1 & $\delta_i+\tau_{i1}$ & 0\\
\hline\hline
\end{tabular*}
\end{center}
\end{table}

Following the design-based approach, we condition on the potential outcomes and treatments $\cF_N=\{D_{i0},D_{i1}(\cdot),Y_{i0}(\cdot),Y_{i1}(\cdot)\}_{i=1}^N$, so that the only source of randomness is the assignment vector $\mathbf Z=(Z_1,\dots,Z_N)'$. We adopt the assignment model of \citet{RambachanRoth2026}.

\begin{assumption}[Assignment design]\label{ass:design}
For some fixed $N_1\in\{1,\dots,N-1\}$ and unknown $p_i\in(0,1)$, possibly dependent on $\cF_N$,
\[
\Prob(\mathbf Z=\mathbf z\mid \cF_N)\;\propto\;\prod_{i=1}^N p_i^{z_i}(1-p_i)^{1-z_i}\quad\text{if }\textstyle\sum_i z_i=N_1,\qquad \text{and } 0 \text{ otherwise.}
\]
\end{assumption}

Assumption~\ref{ass:design} describes independent Bernoulli assignment conditional on the number of assigned units, also known as rejective or conditional Poisson sampling \citep{Hajek1964,ChenDempsterLiu1994}. We write $\PR$, $\ER$ and $\VR$ for probabilities, expectations and variances over this distribution, $N_0=N-N_1$, $p_N=N_1/N$, and $\pi_i=\PR(Z_i=1)$ for the marginal assignment probability, so that $\sum_i\pi_i=N_1$. Because the $p_i$ may depend on potential outcomes and potential take-up, the design allows rich forms of selection into assignment. Complete randomization is the special case $p_i\equiv p$, which implies $\pi_i\equiv p_N$. Finite-population covariances are denoted $\CovN(x_i,y_i)=N^{-1}\sum_i(x_i-\bar x)(y_i-\bar y)$.

\subsection{Estimators and the design-based estimand}

For any variable $X_i=X_i(Z_i)$, let $\hatt_X=N_1^{-1}\sum_i Z_iX_i-N_0^{-1}\sum_i(1-Z_i)X_i$ denote the difference in means. The reduced-form and first-stage DiD estimators are $\hatt_\RF=\hatt_{\Delta Y}$ and $\hatt_\FS=\hatt_{\Delta D}$, and the Wald-DiD estimator is $\widehat\beta=\hatt_\RF/\hatt_\FS$. In a balanced two-period panel, $\hatt_\RF$ and $\hatt_\FS$ coincide with the two-way fixed effects DiD coefficients, and $\widehat\beta$ coincides with the IV estimator in a regression of $\Delta Y_i$ on $\Delta D_i$ instrumented by $Z_i$. Define the weights
\begin{equation}\label{eq:weights}
b_i=\frac{\pi_i}{N_1},\qquad a_i=\frac{\pi_i}{N_1}-\frac{1-\pi_i}{N_0}=\frac{\pi_i-p_N}{N\,p_N(1-p_N)},
\end{equation}
which satisfy $\sum_ib_i=1$ and $\sum_ia_i=0$. Let $\tau_\RF=\ER[\hatt_\RF]$ and $\tau_\FS=\ER[\hatt_\FS]$. Following \citet{RambachanRoth2026}, the \emph{Wald-DiD design estimand} is
\begin{equation}\label{eq:betaR}
\beta_R=\frac{\tau_\RF}{\tau_\FS}\qquad\text{whenever }\tau_\FS\neq0.
\end{equation}
The parameter $\beta_R$ is well defined given $\cF_N$ and the design, and $\tau_\RF$ and $\tau_\FS$ are unbiasedly estimable. The question is what $\beta_R$ means. The natural causal benchmark is the assignment-weighted complier effect
\begin{equation}\label{eq:betaC}
\beta_C=\frac{\sum_{i\in\cC}\pi_i\tau_{i1}}{\sum_{i\in\cC}\pi_i}=\frac{\ER\bigl[\sum_{i\in\cC}Z_i\tau_{i1}\bigr]}{\ER\bigl[\sum_{i\in\cC}Z_i\bigr]},
\end{equation}
the expected average effect among \emph{assigned} compliers. It is the design analogue of a local average treatment effect on the treated (the $\mathrm{LATE}_\pi$ of \citealp{RambachanRoth2026}). Under complete randomization it reduces to the complier average effect $\mathrm{LATE}_\cC=N_\cC^{-1}\sum_{i\in\cC}\tau_{i1}$. Throughout we assume $N_\cC\ge1$.

\section{Identification}\label{sec:ident}

We begin with a general lemma that requires neither monotonicity nor exclusion.

\begin{lemma}[Design expectation of a difference in means]\label{lem:dim}
Under Assumption~\ref{ass:design}, for any fixed potential responses $\{X_i(0),X_i(1)\}_{i=1}^N$,
\[
\ER[\hatt_X]=\sum_{i=1}^N b_i\bigl\{X_i(1)-X_i(0)\bigr\}+\sum_{i=1}^N a_iX_i(0),\qquad
\sum_{i=1}^N a_iX_i(0)=\frac{\CovN\bigl(\pi_i,X_i(0)\bigr)}{p_N(1-p_N)}.
\]
\end{lemma}

The first term is a $\pi$-weighted average effect of assignment on $X$. The second is a selection term that vanishes when assignment probabilities are uncorrelated with the response without assignment. Lemma~\ref{lem:dim}, and hence Theorem~\ref{thm:decomp} below, uses Assumption~\ref{ass:design} only through the facts that $\sum_iZ_i=N_1$ is fixed and that $\pi_i\in(0,1)$; both results therefore hold for any such design. Applying Lemma~\ref{lem:dim} to $X_i=\Delta D_i$ and $X_i=\Delta Y_i$ and using Table~\ref{tab:types} yields our main result. Define the \emph{design contamination} terms
\begin{equation}\label{eq:RF-contam}
\begin{aligned}
R_N&=\sum_{i=1}^Na_i\,\Delta Y_i(D_{i}(0))=\frac{\CovN(\pi_i,\Delta Y_i(D_{i}(0)))}{p_N(1-p_N)},\\
F_N&=\sum_{i=1}^Na_i\,\Delta D_i(0)=\frac{\CovN(\pi_i,\one\{i\in\cS\})}{p_N(1-p_N)},
\end{aligned}
\end{equation}
and the complier mass $B_\cC=\sum_{i\in\cC}b_i>0$.

\begin{theorem}[Decomposition of the Wald-DiD design estimand]\label{thm:decomp}
Suppose Assumptions~\ref{ass:exclusion}--\ref{ass:design} hold. (a) The first stage and reduced form satisfy $\tau_\FS=B_\cC+F_N$ and $\tau_\RF=B_\cC\beta_C+R_N$, where
\begin{align}
F_N&=\sum_{i\in\cS}a_i, \label{eq:FS-decomp}\\
R_N&=\sum_{i\in\cS}a_i\tau_{i1}+\sum_{i\in\cA}a_i\Delta\tau_i+\sum_{i=1}^Na_i\delta_i. \label{eq:RF-decomp}
\end{align}
(b) If $\tau_\FS\neq0$,
\begin{equation}\label{eq:bias}
\beta_R-\beta_C=\frac{R_N-\beta_CF_N}{\tau_\FS}
=\frac{\sum_{i\in\cS}a_i(\tau_{i1}-\beta_C)+\sum_{i\in\cA}a_i\Delta\tau_i+\sum_{i=1}^Na_i\delta_i}{\sum_{i\in\cC}b_i+\sum_{i\in\cS}a_i}.
\end{equation}
(c) Let $\Gamma_i=\Delta Y_i(D_{i}(0))-\beta_C\,\Delta D_i(0)=\delta_i+\one\{i\in\cA\}\Delta\tau_i+\one\{i\in\cS\}(\tau_{i1}-\beta_C)$. If $\tau_\FS\neq0$, $\beta_R=\beta_C$ if and only if $\CovN(\pi_i,\Gamma_i)=0$.
\end{theorem}

Equation~\eqref{eq:bias} isolates three channels through which $\beta_R$ departs from a causal parameter, which are the channels that \citet{deChaisemartinDHaultfoeuille2018} identify in the sampling-based framework. Switchers contribute the deviation of their effects from $\beta_C$, weighted by $a_i$. Always-takers contribute the change in their treatment effects between periods. Trends in untreated outcomes contribute whenever their covariance with the assignment probabilities is different from zero, which is a design-based failure of parallel trends. In the design-based framework, all three channels operate only through heterogeneity in $\pi_i$.

\begin{corollary}[Complete randomization]\label{cor:cr}
If $\pi_i=p_N$ for all $i$, then $a_i=0$, $\tau_\FS=N_\cC/N$ and $\beta_R=\mathrm{LATE}_\cC$, with no restriction on switchers, always-takers or trends.
\end{corollary}

\begin{corollary}[Sufficient conditions]\label{cor:sufficient}
Under the conditions of Theorem~\ref{thm:decomp}, $\beta_R=\beta_C$ if (a) $\CovN(\pi_i,\delta_i)=0$; (b) $\CovN(\pi_i,\one\{i\in\cA\}\Delta\tau_i)=0$; and (c) $\CovN(\pi_i,\one\{i\in\cS\}(\tau_{i1}-\beta_C))=0$. In particular, (a)--(c) hold if $\CovN(\pi_i,\delta_i)=0$, $\tau_{i1}$ is constant and $\Delta\tau_i=0$ for $i\in\cA$, even if $F_N\neq0$.
\end{corollary}

Condition (a) is design-based parallel trends. Condition (b) holds, for example, if effects on always-takers are stable over time, and condition (c) if switchers' effects equal $\beta_C$. The last statement of Corollary~\ref{cor:sufficient} shows that switchers may be concentrated among units with high assignment probabilities. Conditions (a)--(c) of Corollary~\ref{cor:sufficient} are design-based counterparts of the parallel trends, stable treatment effect and homogeneous treatment effect assumptions of \citet{deChaisemartinDHaultfoeuille2018}. Unlike in their framework, none of them is needed under complete randomization (Corollary~\ref{cor:cr})

\begin{remark}[Relation to existing conditions]\label{rem:RR}
\textnormal{Applied to first-differenced data, the IV results of \citet[][Section 5.3]{RambachanRoth2026} give $\beta_R=\beta_C$ under $R_N=F_N=0$, that is, design-based parallel trends in both potential outcomes and potential treatments. Theorem~\ref{thm:decomp}(c) shows that the sharp requirement is the single restriction $R_N=\beta_CF_N$, since $\CovN(\pi_i,\Gamma_i)=p_N(1-p_N)(R_N-\beta_CF_N)$.}
\end{remark}

\begin{remark}[Negative weights]\label{rem:weights}
\textnormal{Theorem~\ref{thm:decomp} can be written as
\[
\beta_R=\sum_{i\in\cC\cup\cS}\omega_i\tau_{i1}+\tau_\FS^{-1}\Bigl(\sum_{i\in\cA}a_i\Delta\tau_i+\sum_ia_i\delta_i\Bigr),
\]
with $\omega_i=b_i/\tau_\FS$ for compliers and $\omega_i=a_i/\tau_\FS$ for switchers. The $\omega_i$ sum to one, but switchers with $\pi_i<p_N$ receive negative weight. Hence, even under design-based parallel trends, $\beta_R$ can be negative when every $\tau_{i1}$ is positive. This parallels the negative-weights results of \citet{deChaisemartinDHaultfoeuille2018,deChaisemartinDHaultfoeuille2020}.}
\end{remark}

\begin{remark}[Testable implications and always-takers]\label{rem:placebo}
\textnormal{Because $D_{i0}$ and $Y_{i0}$ are predetermined, Lemma~\ref{lem:dim} implies that $\ER[\hatt_W]=\CovN(\pi_i,W_i)/\{p_N(1-p_N)\}$ for any predetermined $W_i$. Balance tests on $D_{i0}$ (that is, on $\one\{i\in\cA\}$), on $Y_{i0}$ or on pre-period trends are therefore informative about the covariance of $\pi_i$ with the components of $\Gamma_i$, and Theorem~\ref{thm:inference} below delivers valid inference for them. Moreover, since always-takers are observed at baseline, one can drop units with $D_{i0}=1$. Conditional on the number of assigned units in the remaining subpopulation, the assignment vector there is again of the form in Assumption~\ref{ass:design}, and the term $\sum_{i\in\cA}a_i\Delta\tau_i$ disappears.}
\end{remark}

\section{Inference}\label{sec:inference}

Theorem~\ref{thm:decomp} shows that interpreting $\beta_R$ as $\beta_C$ requires a covariance restriction that the design itself rarely justifies. We therefore develop inference for $\beta_R$, which requires no such restriction, and return to the complier effect in Section~\ref{sec:sensitivity}. Assumption~\ref{ass:monotone} is not used in this section.

\paragraph{Test statistic}
For a hypothesized value $b\in\mathbb R$, let $R_i(z;b)=\Delta Y_i(z)-b\,\Delta D_i(z)$ and $R_i(b)=\Delta Y_i-b\,\Delta D_i$. Since the difference in means is linear,
\[
\widehat M(b)=\hatt_\RF-b\,\hatt_\FS=\hatt_{R(b)},\qquad \ER[\widehat M(b)]=m_N(b):=\tau_\RF-b\,\tau_\FS.
\]
When $\tau_\FS\ne0$, $H_0:\beta_R=b$ is equivalent to $m_N(b)=0$. Let $\hat s_{XW,1}$ and $\hat s_{XW,0}$ be the sample covariances of $(X_i,W_i)$ among assigned and unassigned units, with denominators $N_1-1$ and $N_0-1$, and $\hat v_{XW}=\hat s_{XW,1}/N_1+\hat s_{XW,0}/N_0$. The Neyman-type variance estimator of $\widehat M(b)$ is
\begin{equation}\label{eq:Vhat}
\widehat V(b)=\hat v_{YY}-2b\,\hat v_{YD}+b^2\hat v_{DD},
\end{equation}
where $Y$ and $D$ stand for $\Delta Y$ and $\Delta D$. Equivalently, $\widehat V(b)$ is the HC2 variance of the coefficient on $Z_i$ in an ordinary least squares regression of $R_i(b)$ on $(1,Z_i)$ \citep{ImbensKolesar2016}. The Anderson--Rubin-type statistic is $T_N(b)=\widehat M(b)/\widehat V(b)^{1/2}$.

\paragraph{Asymptotic framework}
We consider a sequence of finite populations indexed by $N$, with all quantities implicitly depending on $N$. Let $c_i(b)=R_i(1;b)/N_1+R_i(0;b)/N_0$, $\tilde\pi_i=\pi_i(1-\pi_i)$, $\bar c(b)=\sum_i\tilde\pi_ic_i(b)/\sum_i\tilde\pi_i$ and
\begin{equation}\label{eq:VN}
V_N(b)=\sum_{i=1}^N\tilde\pi_i\bigl\{c_i(b)-\bar c(b)\bigr\}^2.
\end{equation}

\begin{assumption}[Regularity]\label{ass:regularity}
For a given $b$, (a) there is $\eta\in(0,1/2)$ such that $\eta\le\pi_i\le1-\eta$ for all $i$ and $N$, and $p_N\to p\in(0,1)$; (b) $\sup_N N^{-1}\sum_{i}\{\Delta Y_i(z)^4+\Delta D_i(z)^4\}<\infty$ for $z\in\{0,1\}$; (c) $\liminf_N N\,V_N(b)>0$.
\end{assumption}

Part (a) requires overlap. Part (b) is a standard moment bound for finite-population limit theorems \citep{LiDing2017}. Part (c) rules out degenerate cases in which $\widehat M(b)$ has negligible design variance. Define $m_1(b)=N_1^{-1}\sum_i\pi_iR_i(1;b)$, $m_0(b)=N_0^{-1}\sum_i(1-\pi_i)R_i(0;b)$, $e_i(z;b)=R_i(z;b)-m_z(b)$ and
\begin{equation}\label{eq:Vplus}
V_N^+(b)=\sum_{i=1}^N\frac{\pi_i\,e_i(1;b)^2}{N_1^2}+\sum_{i=1}^N\frac{(1-\pi_i)\,e_i(0;b)^2}{N_0^2}.
\end{equation}

\begin{theorem}[Design-based Anderson--Rubin inference]\label{thm:inference}
Suppose Assumptions~\ref{ass:exclusion}, \ref{ass:design} and \ref{ass:regularity} hold for a given $b$. (a) $\widehat M(b)-m_N(b)=\sum_i(Z_i-\pi_i)c_i(b)$, $\VR[\widehat M(b)]=V_N(b)\{1+o(1)\}$, and $\{\widehat M(b)-m_N(b)\}/\VR[\widehat M(b)]^{1/2}\dto N(0,1)$. (b) $N\{\widehat V(b)-V_N^+(b)\}\pto0$, and
\begin{equation}\label{eq:conservative}
V_N^+(b)-V_N(b)\;\ge\;\sum_{i=1}^N\Bigl\{\frac{\pi_i\,e_i(1;b)}{N_1}-\frac{(1-\pi_i)\,e_i(0;b)}{N_0}\Bigr\}^2\;\ge\;0.
\end{equation}
(c) If $m_N(b)=0$ for all $N$, then $\limsup_{N\to\infty}\PR\{|T_N(b)|>z_{1-\alpha/2}\}\le\alpha$ for every $\alpha\in(0,1)$.
\end{theorem}

Part (c) does not involve $\tau_\FS$. The test therefore remains valid when the first stage is weak, for instance along sequences with $\tau_\FS=h/\sqrt N$, under which Wald inference fails \citep{StaigerStock1997}. Part (b) shows that the variance estimator is conservative even when the effects of assignment on $R_i(b)$ are constant, unless $\pi_i$ is constant. This mirrors \citet[][Proposition 3.2]{RambachanRoth2026} for the difference in means. Sharper but more complex variance bounds in the spirit of \citet{AronowGreenLee2014} could be developed.

\begin{corollary}[Anderson--Rubin confidence set]\label{cor:arset}
Let $\mathrm{CS}^{\AR}_\alpha=\{b:\widehat M(b)^2\le z^2_{1-\alpha/2}\widehat V(b)\}$. If $\tau_\FS\ne0$ and Assumption~\ref{ass:regularity} holds at $b=\beta_R$, $\liminf_N\PR(\beta_R\in\mathrm{CS}^{\AR}_\alpha)\ge1-\alpha$. With $\kappa=z^2_{1-\alpha/2}$, $A=\hatt_\FS^2-\kappa\hat v_{DD}$, $B=\hatt_\RF\hatt_\FS-\kappa\hat v_{YD}$ and $C=\hatt_\RF^2-\kappa\hat v_{YY}$, $\mathrm{CS}^{\AR}_\alpha=\{b:Ab^2-2Bb+C\le0\}$, which is (a) the bounded interval $[(B-\sqrt{B^2-AC})/A,\,(B+\sqrt{B^2-AC})/A]$, containing $\widehat\beta$, if $A>0$; (b) the union of two disjoint half-lines if $A<0$ and $B^2>AC$; (c) the real line if $A<0$ and $B^2<AC$. Case (a) occurs if and only if the first-stage $t$-test rejects $\tau_\FS=0$ at level $\alpha$.
\end{corollary}

\begin{corollary}[Wald inference under a strong first stage]\label{cor:wald}
If Assumption~\ref{ass:regularity} holds at $b=\beta_R$ and $\liminf_N|\tau_\FS|>0$, then, with $\widehat{\mathrm{se}}=\widehat V(\widehat\beta)^{1/2}/|\hatt_\FS|$, $\limsup_N\PR(|\widehat\beta-\beta_R|>z_{1-\alpha/2}\widehat{\mathrm{se}})\le\alpha$.
\end{corollary}

\begin{remark}[Fixed populations and exact tests]
\textnormal{The null $m_N(b)=0$ is not sharp, so Fisher randomization tests are not directly available without further restrictions \citep{ImbensRosenbaum2005,KangPeckKeele2018}. In addition, unlike in a randomized experiment, the researcher does not know the $\pi_i$. Berry--Esseen bounds for rejective sampling, as used by \citet{RambachanRoth2026}, can replace the sequence-of-populations argument.}
\end{remark}

\section{Sensitivity analysis and known designs}\label{sec:sensitivity}

\subsection{Bounded contamination}

Theorem~\ref{thm:decomp}(a) gives the identity
\begin{equation}\label{eq:betaC-identity}
\beta_C=\frac{\tau_\RF-R_N}{\tau_\FS-F_N},
\end{equation}
so the complier effect is determined by the identified $(\tau_\RF,\tau_\FS)$ and the two contamination terms in \eqref{eq:RF-contam}. These terms have direct interpretations: $R_N$ is the covariance between assignment probabilities and the outcome changes units would experience without the policy, and $F_N$ is the covariance between assignment probabilities and switching. We bound them.

\begin{assumption}[Bounded design contamination]\label{ass:bounds}
$|R_N|\le M_R$ and $|F_N|\le M_F$ for known $M_R,M_F\ge0$.
\end{assumption}

\begin{proposition}[Identified set]\label{prop:idset}
Under Assumptions~\ref{ass:exclusion}--\ref{ass:design} and \ref{ass:bounds}, $\beta_C\in\Theta_I=\{(\tau_\RF-r)/(\tau_\FS-f):|r|\le M_R,\ |f|\le M_F,\ \tau_\FS-f>0\}$. If $M_F<\tau_\FS$, $\Theta_I=[\beta^{\mathrm{lo}},\beta^{\mathrm{hi}}]$ with
\[
\beta^{\mathrm{lo}}=\min_{s\in\{-1,1\}}\frac{\tau_\RF-M_R}{\tau_\FS+sM_F},\qquad
\beta^{\mathrm{hi}}=\max_{s\in\{-1,1\}}\frac{\tau_\RF+M_R}{\tau_\FS+sM_F}.
\]
\end{proposition}

\begin{theorem}[Sensitivity confidence set]\label{thm:sens}
Let
\[
\mathrm{CS}^{S}_\alpha(M_R,M_F)=\Bigl\{b:\ |\widehat M(b)|\le M_R+|b|\,M_F+z_{1-\alpha/2}\widehat V(b)^{1/2}\Bigr\}.
\]
Under Assumptions~\ref{ass:exclusion}--\ref{ass:design} and \ref{ass:bounds}, and Assumption~\ref{ass:regularity} at $b=\beta_C$, $\liminf_N\PR\{\beta_C\in\mathrm{CS}^{S}_\alpha(M_R,M_F)\}\ge1-\alpha$. For $M_R=M_F=0$, $\mathrm{CS}^S_\alpha$ equals $\mathrm{CS}^\AR_\alpha$.
\end{theorem}

The confidence set requires no lower bound on the first stage and is computed by a one-dimensional grid search. Because it uses a two-sided critical value at every $b$, it is conservative when $\Theta_I$ is wide relative to sampling uncertainty; refinements in the spirit of \citet{ImbensManski2004} and \citet{Stoye2009} are possible.

\begin{corollary}[Robustness of the null of no effect]\label{cor:breakdown}
$0\in\mathrm{CS}^S_\alpha(M_R,M_F)$ if and only if $|\hatt_\RF|\le M_R+z_{1-\alpha/2}\hat v_{YY}^{1/2}$, whatever the value of $M_F$. Hence the breakdown value for rejecting $\beta_C=0$ is $M_R^\ast=\max\{|\hatt_\RF|-z_{1-\alpha/2}\hat v_{YY}^{1/2},\,0\}$, or $p_N(1-p_N)M_R^\ast$ in covariance units.
\end{corollary}

Corollary~\ref{cor:breakdown} has a simple practical message. Conclusions about the \emph{existence} and sign of a complier effect depend only on how much the assignment probabilities may covary with outcome changes in the absence of the policy. Conclusions about its \emph{magnitude} also depend on how they covary with switching.

\begin{remark}[Calibration]\label{rem:calibration}
\textnormal{By the Cauchy--Schwarz inequality,
\[
|R_N|\le\frac{\sigma_\pi\,\mathrm{SD}_N(\Delta Y_i(0))}{p_N(1-p_N)},\qquad |F_N|\le\frac{\sigma_\pi\{s_\cS(1-s_\cS)\}^{1/2}}{p_N(1-p_N)},
\]
where $\sigma_\pi=\mathrm{SD}_N(\pi_i)$ measures the departure from complete randomization and $s_\cS$ is the switcher share. The $(1-\pi_i)$-weighted analogues of $\mathrm{SD}_N(\Delta Y_i(0))$ and $s_\cS$ are estimable from unassigned units. This reduces the analysis to a single interpretable parameter $\sigma_\pi$, possibly scaled by assumed correlations. Alternatively, following \citet{RambachanRoth2023}, the covariance of $\pi_i$ with pre-period trends, which is estimable by Remark~\ref{rem:placebo}, can benchmark $\CovN(\pi_i,\delta_i)$, and the decomposition \eqref{eq:RF-decomp} allows each channel to be bounded separately.}
\end{remark}

\begin{remark}[Relaxing monotonicity and exclusion]\label{rem:relax}
\textnormal{Lemma~\ref{lem:dim} does not use Assumption~\ref{ass:monotone}, and Theorems~\ref{thm:inference} and \ref{thm:sens} use only identity~\eqref{eq:betaC-identity}. Time monotonicity is needed only to interpret $R_N$ and $F_N$ through compliance types; units that stop treatment enter $R_N$ and $F_N$ and require no change. If assignment monotonicity fails, with defiers $\mathcal D$ and $B_{\mathcal D}=\sum_{i\in\mathcal D}b_i$, then \eqref{eq:betaC-identity} holds with $R_N$ replaced by $R_N-\sum_{i\in\mathcal D}b_i\tau_{i1}$ and $F_N$ by $F_N-B_{\mathcal D}$. Bounds $B_{\mathcal D}\le\bar B$ and $|\tau_{i1}|\le\bar\tau$ on $\mathcal D$ thus enlarge $(M_R,M_F)$ to $(M_R+\bar B\bar\tau,M_F+\bar B)$ (cf.\ \citealp{deChaisemartin2017}). Similarly, a direct effect $\gamma_i$ of assignment on period-$1$ outcomes with $|\gamma_i|\le\bar\gamma$ adds $\sum_ib_i\gamma_i\in[-\bar\gamma,\bar\gamma]$ to $\tau_\RF$, since $\sum_ib_i=1$, and enlarges $M_R$ to $M_R+\bar\gamma$.}
\end{remark}

\subsection{Known assignment probabilities}\label{sec:known}

In stratified experiments and in lotteries with known odds, the $\pi_i$ are known, and the contamination terms are no longer nuisance parameters to be bounded but estimable quantities. We state the result for stratified designs, which covers the application in Section~\ref{sec:app}.

\begin{assumption}[Stratified assignment]\label{ass:strata}
The units are partitioned into fixed strata $\mathcal G_1,\dots,\mathcal G_G$ of sizes $N_g$, and $\mathbf Z$ is completely randomized within each stratum, independently across strata, with fixed $N_{1g}\in\{1,\dots,N_g-1\}$ assigned units in stratum $g$.
\end{assumption}

Under Assumption~\ref{ass:strata}, $\pi_i=\pi_g:=N_{1g}/N_g$ for $i\in\mathcal G_g$, and $N_1=\sum_gN_{1g}$ is fixed, so Lemma~\ref{lem:dim} and Theorem~\ref{thm:decomp} apply. Let $\bar X_{zg}$ be the mean of $X_i$ among units in stratum $g$ with $Z_i=z$, $N_{0g}=N_g-N_{1g}$, and define the stratified and contamination estimators
\begin{equation}\label{eq:strat}
\hat\theta_X=\sum_{g=1}^G\frac{N_{1g}}{N_1}\bigl(\bar X_{1g}-\bar X_{0g}\bigr),\qquad
\hat\rho_X=\sum_{g=1}^G\Bigl(\frac{N_{1g}}{N_1}-\frac{N_{0g}}{N_0}\Bigr)\bar X_{0g},
\end{equation}
which satisfy $\hatt_X=\hat\theta_X+\hat\rho_X$ identically. Let $\widetilde M(b)=\hat\theta_{\Delta Y}-b\,\hat\theta_{\Delta D}$, and let $\widetilde V(b)=\sum_g(N_{1g}/N_1)^2\{\hat s^2_{1g}(b)/N_{1g}+\hat s^2_{0g}(b)/N_{0g}\}$, where $\hat s^2_{zg}(b)$ is the sample variance of $R_i(b)$ among units in stratum $g$ with $Z_i=z$.

\begin{proposition}[Known assignment probabilities]\label{prop:known}
Suppose Assumptions~\ref{ass:exclusion}, \ref{ass:monotone} and \ref{ass:strata} hold. (a) $\ER[\hat\theta_{\Delta D}]=B_\cC$, $\ER[\hat\theta_{\Delta Y}]=B_\cC\beta_C$, $\ER[\hat\rho_{\Delta D}]=F_N$ and $\ER[\hat\rho_{\Delta Y}]=R_N$; hence $\beta_C=\ER[\hat\theta_{\Delta Y}]/\ER[\hat\theta_{\Delta D}]$ and $\beta_R-\beta_C=(R_N-\beta_CF_N)/\tau_\FS$ are identified. (b) Suppose further that $G$ is fixed, that $\eta\le\pi_g\le1-\eta$ for some $\eta>0$ and $N_g/N\to\lambda_g>0$, that the moment bound of Assumption~\ref{ass:regularity}(b) holds within each stratum, and that $\liminf_N N\,\VR[\widetilde M(b)]>0$. Under $H_0:\beta_C=b$, $\limsup_N\PR\{|\widetilde M(b)|>z_{1-\alpha/2}\widetilde V(b)^{1/2}\}\le\alpha$.
\end{proposition}

Part (a) holds without any restriction on trends, switchers or always-takers: with a known design, randomization alone identifies $\beta_C$. The same argument applied to period-$1$ levels shows that the stratified IV estimator in levels has the same estimand, because $\ER[\hat\theta_{Y_0}]=\ER[\hat\theta_{D_0}]=0$ for predetermined variables. The role of differencing is then only to improve precision, as in covariate adjustment. Part (b) gives a stratified Anderson--Rubin test that is again robust to a weak first stage. The estimators $\hat\rho_{\Delta Y}$ and $\hat\rho_{\Delta D}$ measure the contamination that the pooled Wald-DiD estimator incurs by ignoring the heterogeneity in $\pi_i$; they depend only on unassigned units, whose outcome changes reveal $\Delta Y_i(D_{i}(0))$ and $\Delta D_i(0)$.

\section{Simulation evidence}\label{sec:sim}

We draw one finite population per design, hold its potential outcomes and potential treatment decisions fixed, and redraw only the assignment, $5{,}000$ times, from Assumption~\ref{ass:design} with $N_1=N/2$ and $p_i=\mathrm{logit}^{-1}(c+\gamma W_i)$ for a unit characteristic $W_i\sim N(0,1)$. Compliance-type shares are $(0.40,0.10,0.35,0.15)$ for $(\cN,\cA,\cC,\cS)$, and effects and trends depend on $W_i$. In D1 assignment is completely randomized ($\gamma=0$). In D2 assignment probabilities are correlated with untreated trends. In D3 switchers are concentrated among high-$\pi_i$ units, with design-based parallel trends and homogeneous effects. D4 is as D3 but with heterogeneous effects. D5 has a weak first stage ($N=200$, $4\%$ compliers) under complete randomization. Section S2 of the Online Appendix gives the details and additional results.

\begin{table}[t]
\caption{\label{tab:sim}Design-based Monte Carlo results.}
\begin{center}
\small
\begin{tabular*}{\textwidth}{@{\extracolsep{\fill}}lrrrrrrrrrr@{}}
\hline\hline
 & $N$ & $\beta_R$ & $\beta_C$ & $\tau_\FS$ & Bias & \multicolumn{2}{c}{Coverage of $\beta_R$} & \multicolumn{2}{c}{Coverage of $\beta_C$} & $\overline{V}/\VR$\\
 & & & & & & Wald & AR & AR & Sens. & \\
\hline
D1 & 1000 & 0.989 & 0.989 & 0.349 & 0.000 & 0.962 & 0.960 & 0.960 & 0.960 & 1.05\\
D2 & 1000 & 2.561 & 1.254 & 0.377 & 0.006 & 0.985 & 0.985 & 0.000 & 0.997 & 1.57\\
D3 & 1000 & 1.000 & 1.000 & 0.412 & $-$0.001 & 0.969 & 0.968 & 0.968 & 1.000 & 1.21\\
D4 & 1000 & 1.539 & 1.195 & 0.405 & $-$0.002 & 0.975 & 0.974 & 0.123 & 1.000 & 1.34\\
D5 & 200 & 1.034 & 1.079 & 0.020 & 2.288 & 0.782 & 0.951 & 0.951 & 0.952 & 1.01\\
\hline\hline
\end{tabular*}
\end{center}
\footnotesize
\renewcommand{\baselineskip}{11pt}
\textbf{Note:} Results over $5{,}000$ assignment draws per design. Bias is the median of $\widehat\beta-\beta_R$. Wald is the interval of Corollary~\ref{cor:wald}, AR is $\mathrm{CS}^\AR_{0.05}$ and Sens.\ is $\mathrm{CS}^S_{0.05}(|R_N|,|F_N|)$, with bounds set at the true contamination. The last column is the mean of $\widehat V(\beta_R)$ divided by the simulation variance of $\widehat M(\beta_R)$. In D5, $\mathrm{CS}^\AR_{0.05}$ is unbounded in $94\%$ of draws.
\end{table}

Table~\ref{tab:sim} confirms the theory. In every design the AR set covers the design estimand $\beta_R$ at or above the nominal rate, up to simulation error. The variance estimator is conservative, more so when assignment probabilities are heterogeneous (D2--D4), as predicted by \eqref{eq:conservative}. Under complete randomization, $\beta_R$ equals the complier effect (Corollary~\ref{cor:cr}). When assignment probabilities covary with untreated trends (D2), or with switching when effects are heterogeneous (D4), $\beta_R$ departs substantially from $\beta_C$, and inference that treats $\beta_R$ as a complier effect fails. In D3, switchers are concentrated among likely-assigned units ($F_N=0.071$), yet $\beta_R=\beta_C$ because $R_N=\beta_CF_N$, which illustrates that the condition in Theorem~\ref{thm:decomp}(c) is strictly weaker than $R_N=F_N=0$. The sensitivity set with correctly specified bounds restores coverage of $\beta_C$. With a weak first stage (D5), the Wald interval undercovers ($78\%$) and the estimator is badly median biased, whereas the AR set retains nominal coverage.

\section{Empirical application: the Oregon Health Insurance Experiment}\label{sec:app}

\subsection{Setting and fit with the framework}

In 2008 the state of Oregon reopened its Medicaid programme for low-income uninsured adults (OHP Standard) and allocated the right to apply by lottery among the individuals on a waiting list \citep{Finkelstein2012}. Eight drawings took place between March and September 2008. Selected individuals, together with the members of their household on the list, could apply, and about 30\% of those selected enrolled. \citet{Taubman2014} linked the list to emergency-department (ED) records from the Portland-area hospitals covering 1 January 2007 to 30 September 2009. We use their individual-level public-use data \citep{OHIEdata}.

The mapping to our framework is direct. The units are the individuals in the ED sample, $Z_i$ indicates selection in any drawing, and $Y_{it}$ is the number of ED visits per year in period $t$, where period $0$ runs from 1 January 2007 to 9 March 2008 and period $1$ from 10 March 2008, the earliest notification date, to 30 September 2009. The treatment $D_{i1}$ indicates enrolment in any Medicaid programme during period $1$, the first-stage measure of \citet{Taubman2014}. The public-use files contain no measure of enrolment before the lottery, and we set $D_{i0}=0$. Then there are no always-takers, Assumption~\ref{ass:monotone}(b) holds trivially, and the always-taker channel in \eqref{eq:RF-decomp} is absent.

Three features make the experiment a good fit. First, compliance is imperfect in both directions. Many selected individuals did not enrol, and some individuals who were not selected enrolled through other routes, mostly in OHP Plus. The latter are switchers in Table~\ref{tab:types}, a group that a cross-sectional IV analysis classifies as always-takers. Second, the state drew individuals but extended selection to the whole household, a procedure that favoured larger households \citep{Finkelstein2012}. Assignment probabilities are therefore heterogeneous, which is precisely what generates the contamination terms of Theorem~\ref{thm:decomp}. Third, the ED data form a panel with outcomes measured before randomization.

Unlike in the quasi-experiments for which Sections~\ref{sec:inference} and \ref{sec:sensitivity} are designed, the assignment probabilities are known up to the number of household members on the list ($1$, $2$ or $3$), within which selection is random. The application is therefore a benchmark. Proposition~\ref{prop:known} identifies $\beta_C$, so we can estimate the contamination $(R_N,F_N)$ that an analyst who ignores the heterogeneity in $\pi_i$ would face, compare the pooled Wald-DiD estimate with $\hat\beta_C=\hat\theta_{\Delta Y}/\hat\theta_{\Delta D}$, and compare the estimated contamination with the breakdown value of Section~\ref{sec:sensitivity}. Two features of the lottery require adaptation. Because selection is at the household level, all variances treat households as the units of assignment and are computed from household totals of residuals within each stratum and arm. Because the number of selected individuals was not fixed in advance, we condition on it.

\subsection{Results}

The ED sample contains 24,615 individuals with non-missing outcomes in 22,159 households. The overall selection rate is $p_N=0.390$, and the selection rates for one, two and three household members on the list are 0.344, 0.571 and 0.885, so that $\sigma_\pi=0.094$. Among individuals who were not selected, a share 0.151 enrolled in Medicaid during period $1$; these are the switchers.

Table~\ref{tab:ohie} reports the pooled estimators of Section~\ref{sec:setup}, which ignore the strata and estimate $\tau_\FS$, $\tau_\RF$ and $\beta_R$, and the stratified estimators of Proposition~\ref{prop:known}, which estimate $B_\cC$, $B_\cC\beta_C$ and $\beta_C$. Selection raised Medicaid enrolment by 0.241, so the first stage is strong and the AR sets are bounded intervals. For ED visits per year, the stratified estimate of the complier effect is 0.292 (standard error 0.077), against a period-$1$ mean of 0.64 visits per year among non-selected individuals, consistent with the increase in ED use found by \citet{Taubman2014}. The pooled Wald-DiD estimate is 0.266 (0.077). Their difference, $-0.026$, estimates the bias $(R_N-\beta_CF_N)/\tau_\FS$ of Theorem~\ref{thm:decomp}(b). Both contamination estimates are small: $\hat\rho_{\Delta Y}=-0.007$ (0.005) and $\hat\rho_{\Delta D}=-0.001$ (0.001), so switching is essentially uncorrelated with the assignment probabilities. The breakdown value of Corollary~\ref{cor:breakdown} for the pooled analysis is $M_R^\ast=0.028$, about four times $|\hat\rho_{\Delta Y}|$. An analyst who did not know the design, but was willing to bound the reduced-form contamination at a few hundredths of a visit per year, would therefore have correctly concluded that Medicaid increases ED use among compliers.

\begin{table}[t]
\caption{\label{tab:ohie}Medicaid and emergency-department use in Oregon.}
\begin{center}
\footnotesize
\setlength{\tabcolsep}{3pt}
\begin{tabular*}{\textwidth}{@{\extracolsep{\fill}}lcccccc@{}}
\hline\hline
 & First stage & Reduced form & Wald-DiD & 95\% AR set & $\hat\rho_{\Delta Y}$ & $\hat\rho_{\Delta D}$\\
\hline
\multicolumn{7}{@{}l}{\textit{ED visits per year}}\\
Pooled (ignores strata) & 0.240 & 0.064 & 0.266 & [0.115, 0.417] & $-0.007$ & $-0.001$\\
 & (0.006) & (0.018) & (0.077) & & (0.005) & (0.001)\\
Stratified (known design) & 0.241 & 0.071 & 0.292 & [0.142, 0.443] & & \\
 & (0.006) & (0.019) & (0.077) & & & \\[3pt]
\multicolumn{7}{@{}l}{\textit{Any ED visit}}\\
Pooled (ignores strata) & 0.240 & 0.014 & 0.060 & [0.003, 0.116] & $-0.002$ & $-0.001$\\
 & (0.006) & (0.007) & (0.029) & & (0.002) & (0.001)\\
Stratified (known design) & 0.241 & 0.016 & 0.068 & [0.011, 0.125] & & \\
 & (0.006) & (0.007) & (0.029) & & & \\
\hline\hline
\end{tabular*}
\end{center}
\footnotesize
\renewcommand{\baselineskip}{11pt}
\textbf{Note:} Pooled rows report $\hatt_\FS$, $\hatt_\RF$, $\widehat\beta$ and $\mathrm{CS}^\AR_{0.05}$ (Section~\ref{sec:inference}). Stratified rows report $\hat\theta_{\Delta D}$, $\hat\theta_{\Delta Y}$, $\hat\beta_C$ and the stratified AR set of Proposition~\ref{prop:known}. The last two columns report the contamination estimators $\hat\rho_{\Delta Y}$ and $\hat\rho_{\Delta D}$ in \eqref{eq:strat}. Strata are household size on the lottery list. The treatment is Medicaid enrolment. Standard errors in parentheses treat households as the units of assignment. ED visits per year are visit counts, truncated and censored as in the public-use files, divided by the length of each period in years. Any ED visit compares an indicator for any visit in period $1$ with the same indicator in period $0$, whose windows differ in length. Sample: 24,615 individuals in the ED sample of \citet{Taubman2014}.
\end{table}

Differencing is what makes the pooled analysis robust. Larger households, which were more likely to be selected, used the ED less before the lottery: the pooled difference in pre-period ED visits per year between selected and non-selected individuals is $-0.070$ (0.020), whereas the stratified difference is $-0.011$ (0.020), consistent with successful randomization within strata (Remark~\ref{rem:placebo}). Consequently, the pooled IV estimate in period-$1$ levels is $-0.026$ (0.085), whereas the stratified levels estimate, whose estimand is again $\beta_C$, is 0.245 (0.081). Selection on household size thus enters levels strongly but trends only weakly, and the Wald-DiD estimand inherits only the latter. The any-visit outcome gives the same qualitative picture.
\section{Conclusion}\label{sec:conclusion}

This paper develops a design-based framework for the fuzzy difference-in-differences setting. This approach is motivated by settings in which researchers wish to remain agnostic about the statistical process behind the sampling and formation of potential outcomes and treatments. Viewed through the lens of the assignment mechanism, fuzzy DiD has a transparent structure. The Wald-DiD design estimand equals an assignment-weighted complier effect plus contamination from switchers, always-takers and trends in untreated outcomes, all of which affect the estimand through the covariance between assignment probabilities and a single contamination index. When this covariance cannot be argued to vanish, the design still supports valid, weak-identification-robust and conservative inference on the Wald-DiD estimand, and a sensitivity analysis for the complier effect whose conclusions about the existence of an effect depend only on the reduced-form contamination. When the design is known, the contamination is estimable and the complier effect is identified. Extensions to staggered adoption \citep{AtheyImbens2022,RothSantAnna2023JPE}, clustered assignment, covariate adjustment and alternative fuzzy DiD estimands such as the time-corrected Wald ratio of \citet{deChaisemartinDHaultfoeuille2018} are natural next steps.

\bibliography{refs}

@article{deChaisemartinDHaultfoeuille2018,
  author  = {de Chaisemartin, Cl{\'e}ment and D'Haultf{\oe}uille, Xavier},
  title   = {Fuzzy Differences-in-Differences},
  journal = {Review of Economic Studies},
  year    = {2018},
  volume  = {85},
  pages   = {999--1028},
  doi     = {10.1093/restud/rdx049}
}

@article{deChaisemartinDHaultfoeuille2020,
  author  = {de Chaisemartin, Cl{\'e}ment and D'Haultf{\oe}uille, Xavier},
  title   = {Two-Way Fixed Effects Estimators with Heterogeneous Treatment Effects},
  journal = {American Economic Review},
  year    = {2020},
  volume  = {110},
  number  = {9},
  pages   = {2964--2996},
  doi     = {10.1257/aer.20181169}
}

@article{deChaisemartinDHaultfoeuille2023survey,
  author  = {de Chaisemartin, Cl{\'e}ment and D'Haultf{\oe}uille, Xavier},
  title   = {Two-Way Fixed Effects and Differences-in-Differences with Heterogeneous Treatment Effects: A Survey},
  journal = {Econometrics Journal},
  year    = {2023},
  volume  = {26},
  pages   = {C1--C30},
  doi     = {10.1093/ectj/utac017}
}

@article{CallawaySantAnna2021,
  author  = {Callaway, Brantly and Sant'Anna, Pedro H. C.},
  title   = {Difference-in-Differences with Multiple Time Periods},
  journal = {Journal of Econometrics},
  year    = {2021},
  volume  = {225},
  pages   = {200--230},
  doi     = {10.1016/j.jeconom.2020.12.001}
}

@article{SunAbraham2021,
  author  = {Sun, Liyang and Abraham, Sarah},
  title   = {Estimating Dynamic Treatment Effects in Event Studies with Heterogeneous Treatment Effects},
  journal = {Journal of Econometrics},
  year    = {2021},
  volume  = {225},
  pages   = {175--199},
  doi     = {10.1016/j.jeconom.2020.09.006}
}

@article{GoodmanBacon2021,
  author  = {Goodman-Bacon, Andrew},
  title   = {Difference-in-Differences with Variation in Treatment Timing},
  journal = {Journal of Econometrics},
  year    = {2021},
  volume  = {225},
  pages   = {254--277},
  doi     = {10.1016/j.jeconom.2021.03.014}
}

@article{BorusyakJaravelSpiess2024,
  author  = {Borusyak, Kirill and Jaravel, Xavier and Spiess, Jann},
  title   = {Revisiting Event-Study Designs: Robust and Efficient Estimation},
  journal = {Review of Economic Studies},
  year    = {2024},
  volume  = {91},
  pages   = {3253--3285},
  doi     = {10.1093/restud/rdae007}
}

@article{RambachanRoth2023,
  author  = {Rambachan, Ashesh and Roth, Jonathan},
  title   = {A More Credible Approach to Parallel Trends},
  journal = {Review of Economic Studies},
  year    = {2023},
  volume  = {90},
  pages   = {2555--2591},
  doi     = {10.1093/restud/rdad018}
}

@article{RothSantAnna2023JPE,
  author  = {Roth, Jonathan and Sant'Anna, Pedro H. C.},
  title   = {Efficient Estimation for Staggered Rollout Designs},
  journal = {Journal of Political Economy Microeconomics},
  year    = {2023},
  volume  = {1},
  pages   = {669--709},
  doi     = {10.1086/726581}
}

@article{RothSantAnnaBilinskiPoe2023,
  author  = {Roth, Jonathan and Sant'Anna, Pedro H. C. and Bilinski, Alyssa and Poe, John},
  title   = {What's Trending in Difference-in-Differences? {A} Synthesis of the Recent Econometrics Literature},
  journal = {Journal of Econometrics},
  year    = {2023},
  volume  = {235},
  pages   = {2218--2244},
  doi     = {10.1016/j.jeconom.2023.03.008}
}

@article{Duflo2001,
  author  = {Duflo, Esther},
  title   = {Schooling and Labor Market Consequences of School Construction in {I}ndonesia: Evidence from an Unusual Policy Experiment},
  journal = {American Economic Review},
  year    = {2001},
  volume  = {91},
  number  = {4},
  pages   = {795--813},
  doi     = {10.1257/aer.91.4.795}
}

@article{AtheyImbens2022,
  author  = {Athey, Susan and Imbens, Guido W.},
  title   = {Design-Based Analysis in Difference-in-Differences Settings with Staggered Adoption},
  journal = {Journal of Econometrics},
  year    = {2022},
  volume  = {226},
  pages   = {62--79},
  doi     = {10.1016/j.jeconom.2020.10.012}
}

@article{RambachanRoth2026,
  author  = {Rambachan, Ashesh and Roth, Jonathan},
  title   = {Design-Based Uncertainty for Quasi-Experiments},
  journal = {Journal of the American Statistical Association},
  year    = {2026},
  volume  = {121},
  pages   = {477--491},
  doi     = {10.1080/01621459.2025.2526700}
}

@article{AbadieAtheyImbensWooldridge2020,
  author  = {Abadie, Alberto and Athey, Susan and Imbens, Guido W. and Wooldridge, Jeffrey M.},
  title   = {Sampling-Based versus Design-Based Uncertainty in Regression Analysis},
  journal = {Econometrica},
  year    = {2020},
  volume  = {88},
  pages   = {265--296},
  doi     = {10.3982/ECTA12675}
}

@article{AbadieAtheyImbensWooldridge2023,
  author  = {Abadie, Alberto and Athey, Susan and Imbens, Guido W. and Wooldridge, Jeffrey M.},
  title   = {When Should You Adjust Standard Errors for Clustering?},
  journal = {Quarterly Journal of Economics},
  year    = {2023},
  volume  = {138},
  pages   = {1--35},
  doi     = {10.1093/qje/qjac038}
}

@article{Neyman1990,
  author  = {Splawa-Neyman, Jerzy},
  title   = {On the Application of Probability Theory to Agricultural Experiments. {E}ssay on Principles. {S}ection 9},
  journal = {Statistical Science},
  year    = {1990},
  volume  = {5},
  pages   = {465--472},
  note    = {Translated and edited by D. M. Dabrowska and T. P. Speed from the Polish original (1923)},
  doi     = {10.1214/ss/1177012031}
}

@book{ImbensRubin2015,
  author    = {Imbens, Guido W. and Rubin, Donald B.},
  title     = {Causal Inference for Statistics, Social, and Biomedical Sciences: An Introduction},
  publisher = {Cambridge University Press},
  address   = {Cambridge},
  year      = {2015},
  doi       = {10.1017/CBO9781139025751}
}

@article{LiDing2017,
  author  = {Li, Xinran and Ding, Peng},
  title   = {General Forms of Finite Population Central Limit Theorems with Applications to Causal Inference},
  journal = {Journal of the American Statistical Association},
  year    = {2017},
  volume  = {112},
  pages   = {1759--1769},
  doi     = {10.1080/01621459.2017.1295865}
}

@article{AronowGreenLee2014,
  author  = {Aronow, Peter M. and Green, Donald P. and Lee, Donald K. K.},
  title   = {Sharp Bounds on the Variance in Randomized Experiments},
  journal = {Annals of Statistics},
  year    = {2014},
  volume  = {42},
  pages   = {850--871},
  doi     = {10.1214/13-AOS1200}
}

@article{Hajek1964,
  author  = {H{\'a}jek, Jaroslav},
  title   = {Asymptotic Theory of Rejective Sampling with Varying Probabilities from a Finite Population},
  journal = {Annals of Mathematical Statistics},
  year    = {1964},
  volume  = {35},
  pages   = {1491--1523},
  doi     = {10.1214/aoms/1177700375}
}

@article{ChenDempsterLiu1994,
  author  = {Chen, Xiang-Hui and Dempster, Arthur P. and Liu, Jun S.},
  title   = {Weighted Finite Population Sampling to Maximize Entropy},
  journal = {Biometrika},
  year    = {1994},
  volume  = {81},
  pages   = {457--469},
  doi     = {10.1093/biomet/81.3.457}
}

@article{ImbensKolesar2016,
  author  = {Imbens, Guido W. and Koles{\'a}r, Michal},
  title   = {Robust Standard Errors in Small Samples: Some Practical Advice},
  journal = {Review of Economics and Statistics},
  year    = {2016},
  volume  = {98},
  pages   = {701--712},
  doi     = {10.1162/REST_a_00552}
}

@article{ImbensAngrist1994,
  author  = {Imbens, Guido W. and Angrist, Joshua D.},
  title   = {Identification and Estimation of Local Average Treatment Effects},
  journal = {Econometrica},
  year    = {1994},
  volume  = {62},
  pages   = {467--475},
  doi     = {10.2307/2951620}
}

@article{deChaisemartin2017,
  author  = {de Chaisemartin, Cl{\'e}ment},
  title   = {Tolerating Defiance? {L}ocal Average Treatment Effects without Monotonicity},
  journal = {Quantitative Economics},
  year    = {2017},
  volume  = {8},
  pages   = {367--396},
  doi     = {10.3982/QE601}
}

@article{StaigerStock1997,
  author  = {Staiger, Douglas and Stock, James H.},
  title   = {Instrumental Variables Regression with Weak Instruments},
  journal = {Econometrica},
  year    = {1997},
  volume  = {65},
  pages   = {557--586},
  doi     = {10.2307/2171753}
}

@article{AndrewsStockSun2019,
  author  = {Andrews, Isaiah and Stock, James H. and Sun, Liyang},
  title   = {Weak Instruments in Instrumental Variables Regression: Theory and Practice},
  journal = {Annual Review of Economics},
  year    = {2019},
  volume  = {11},
  pages   = {727--753},
  doi     = {10.1146/annurev-economics-080218-025643}
}

@article{ImbensRosenbaum2005,
  author  = {Imbens, Guido W. and Rosenbaum, Paul R.},
  title   = {Robust, Accurate Confidence Intervals with a Weak Instrument: Quarter of Birth and Education},
  journal = {Journal of the Royal Statistical Society, Series A},
  year    = {2005},
  volume  = {168},
  pages   = {109--126},
  doi     = {10.1111/j.1467-985X.2004.00339.x}
}

@article{KangPeckKeele2018,
  author  = {Kang, Hyunseung and Peck, Laura and Keele, Luke},
  title   = {Inference for Instrumental Variables: A Randomization Inference Approach},
  journal = {Journal of the Royal Statistical Society, Series A},
  year    = {2018},
  volume  = {181},
  pages   = {1231--1254},
  doi     = {10.1111/rssa.12353}
}

@book{Rosenbaum2002,
  author    = {Rosenbaum, Paul R.},
  title     = {Observational Studies},
  edition   = {2nd},
  publisher = {Springer},
  address   = {New York},
  year      = {2002},
  doi       = {10.1007/978-1-4757-3692-2}
}

@article{CinelliHazlett2020,
  author  = {Cinelli, Carlos and Hazlett, Chad},
  title   = {Making Sense of Sensitivity: Extending Omitted Variable Bias},
  journal = {Journal of the Royal Statistical Society, Series B},
  year    = {2020},
  volume  = {82},
  pages   = {39--67},
  doi     = {10.1111/rssb.12348}
}

@article{ManskiPepper2018,
  author  = {Manski, Charles F. and Pepper, John V.},
  title   = {How Do Right-to-Carry Laws Affect Crime Rates? {C}oping with Ambiguity Using Bounded-Variation Assumptions},
  journal = {Review of Economics and Statistics},
  year    = {2018},
  volume  = {100},
  pages   = {232--244},
  doi     = {10.1162/REST_a_00689}
}

@article{ImbensManski2004,
  author  = {Imbens, Guido W. and Manski, Charles F.},
  title   = {Confidence Intervals for Partially Identified Parameters},
  journal = {Econometrica},
  year    = {2004},
  volume  = {72},
  pages   = {1845--1857},
  doi     = {10.1111/j.1468-0262.2004.00555.x}
}

@article{Stoye2009,
  author  = {Stoye, J{\"o}rg},
  title   = {More on Confidence Intervals for Partially Identified Parameters},
  journal = {Econometrica},
  year    = {2009},
  volume  = {77},
  pages   = {1299--1315},
  doi     = {10.3982/ECTA7347}
}

@article{Finkelstein2012,
  author  = {Finkelstein, Amy and Taubman, Sarah and Wright, Bill and Bernstein, Mira and Gruber, Jonathan and Newhouse, Joseph P. and Allen, Heidi and Baicker, Katherine and {Oregon Health Study Group}},
  title   = {The {O}regon Health Insurance Experiment: Evidence from the First Year},
  journal = {Quarterly Journal of Economics},
  year    = {2012},
  volume  = {127},
  pages   = {1057--1106},
  doi     = {10.1093/qje/qjs020}
}

@article{Taubman2014,
  author  = {Taubman, Sarah L. and Allen, Heidi L. and Wright, Bill J. and Baicker, Katherine and Finkelstein, Amy N.},
  title   = {Medicaid Increases Emergency-Department Use: Evidence from {O}regon's Health Insurance Experiment},
  journal = {Science},
  year    = {2014},
  volume  = {343},
  pages   = {263--268},
  doi     = {10.1126/science.1246183}
}

@misc{OHIEdata,
  author       = {Finkelstein, Amy and Baicker, Katherine},
  title        = {The {O}regon Health Insurance Experiment: Data},
  howpublished = {National Bureau of Economic Research. \url{https://doi.org/10.60592/rx0z-sp89}},
  year         = {2020}
}

@techreport{hudson2017interpreting,
  title     = {Interpreting Instrumented Difference-in-Differences},
  author    = {Hudson, Sally and Hull, Peter and Liebersohn, Jack},
  year      = {2017},
  month     = {},
  institution = {Unpublished manuscript},
  type      = {Working Paper}
}

\appendix

\IfFileExists{appendix_proofs.tex}{

\section{Proofs}\label{app:proofs}

Throughout, $p=p_N$ and all sums run over $i=1,\dots,N$ unless indicated otherwise. We use the following properties of the rejective design. Their proof is routine and is given in Section S3 of the Online Appendix.

\begin{lemma}\label{lem:rejective}
Suppose Assumption~\ref{ass:design} holds and let $d_N=\sum_i\tilde\pi_i$. (a) For $i\ne j$, $\Cov_R(Z_i,Z_j)\le0$; consequently, for any real weights $w_i$, $\VR[\sum_iZ_iw_i]\le2\sum_i\tilde\pi_iw_i^2\le\frac12\sum_iw_i^2$. (b) If $d_N\to\infty$, then for any array $\{y_i\}$ with $\bar y=\sum_i\tilde\pi_iy_i/d_N$, $\VR[\sum_iZ_iy_i]=\{1+o(1)\}\sum_i\tilde\pi_i(y_i-\bar y)^2$. (c) If, in addition, $\max_i(y_i-\bar y)^2/\sum_i\tilde\pi_i(y_i-\bar y)^2\to0$, then $\sum_i(Z_i-\pi_i)y_i/\{\sum_i\tilde\pi_i(y_i-\bar y)^2\}^{1/2}\dto N(0,1)$.
\end{lemma}

\subsection{Proofs for Section~\ref{sec:ident}}

\begin{proof}[Proof of Lemma~\ref{lem:dim}]
Since $N_1$ and $N_0$ are fixed, $\ER[\hatt_X]=\sum_i\{\pi_iX_i(1)/N_1-(1-\pi_i)X_i(0)/N_0\}$. Adding and subtracting $\pi_iX_i(0)/N_1$ gives $\sum_ib_i\{X_i(1)-X_i(0)\}+\sum_ia_iX_i(0)$. For the second claim, $a_i=\{\pi_i(N_0+N_1)-N_1\}/(N_1N_0)=(\pi_i-p)/\{Np(1-p)\}$. Since $\sum_i\pi_i=N_1$, the mean of $\pi_i$ is $p$ and $\sum_i(\pi_i-p)X_i(0)=N\CovN(\pi_i,X_i(0))$.
\end{proof}

\begin{proof}[Proof of Theorem~\ref{thm:decomp}]
(a) By Table~\ref{tab:types}, $\Delta D_i(1)-\Delta D_i(0)=\one\{i\in\cC\}$ and $\Delta D_i(0)=\one\{i\in\cS\}$, so Lemma~\ref{lem:dim} gives $\tau_\FS=\sum_{i\in\cC}b_i+\sum_{i\in\cS}a_i=B_\cC+F_N$. Likewise, $\Delta Y_i(1)-\Delta Y_i(0)=\one\{i\in\cC\}\tau_{i1}$ and $\Delta Y_i(0)=\delta_i+\one\{i\in\cA\}\Delta\tau_i+\one\{i\in\cS\}\tau_{i1}$. Hence $\tau_\RF=\sum_{i\in\cC}b_i\tau_{i1}+R_N$ with $R_N$ as in \eqref{eq:RF-decomp}, and $\sum_{i\in\cC}b_i\tau_{i1}=B_\cC\beta_C$ by \eqref{eq:betaC}. Finally, $B_\cC>0$ because $N_\cC\ge1$ and $\pi_i>0$ for all $i$.
(b) From (a), $\tau_\RF-\beta_C\tau_\FS=R_N-\beta_CF_N$. Dividing by $\tau_\FS$ and substituting \eqref{eq:FS-decomp} and \eqref{eq:RF-decomp} gives \eqref{eq:bias}.
(c) $R_N-\beta_CF_N=\sum_ia_i\{\Delta Y_i(0)-\beta_C\Delta D_i(0)\}=\sum_ia_i\Gamma_i=\CovN(\pi_i,\Gamma_i)/\{p(1-p)\}$ by Lemma~\ref{lem:dim}. Since $\tau_\FS\ne0$, part (b) shows that $\beta_R=\beta_C$ if and only if this quantity is zero.
\end{proof}

\begin{proof}[Proof of Corollaries~\ref{cor:cr} and \ref{cor:sufficient}]
If $\pi_i\equiv p$, then $a_i=0$ and $b_i=1/N$, so $R_N=F_N=0$, $B_\cC=N_\cC/N$ and $\beta_C=\mathrm{LATE}_\cC$, and Theorem~\ref{thm:decomp}(a) gives Corollary~\ref{cor:cr}. For Corollary~\ref{cor:sufficient}, $\CovN(\pi_i,\Gamma_i)$ is the sum of the three covariances in (a)--(c) by linearity, so Theorem~\ref{thm:decomp}(c) applies. If $\tau_{i1}=\tau$ for all $i$ and $\Delta\tau_i=0$ for $i\in\cA$, then $\beta_C=\tau$ and (b) and (c) hold trivially, whatever the value of $F_N$.
\end{proof}

\subsection{Proofs for Section~\ref{sec:inference}}

\begin{proof}[Proof of Theorem~\ref{thm:inference}]
Fix $b$ and suppress it from the notation, writing $R_i(z)=R_i(z;b)$ and $e_i(z)=e_i(z;b)$.

\textsc{Step} 1 (exact representation). Since $N_1$ is fixed, $\widehat M=\sum_iZ_iR_i(1)/N_1-\sum_i(1-Z_i)R_i(0)/N_0=\sum_iZ_ic_i-\sum_iR_i(0)/N_0$. Taking expectations, $m_N=\sum_i\pi_ic_i-\sum_iR_i(0)/N_0$, so $\widehat M-m_N=\sum_i(Z_i-\pi_i)c_i$.

\textsc{Step} 2 (asymptotic normality). Assumption~\ref{ass:regularity}(a) gives $d_N\ge N\eta(1-\eta)\to\infty$. By Lemma~\ref{lem:rejective}(b) with $y_i=c_i$, $\VR[\widehat M]=V_N\{1+o(1)\}$. Assumption~\ref{ass:regularity}(b) implies $N^{-1}\sum_iR_i(z)^4\le K$ for some $K<\infty$, and hence $\max_iR_i(z)^2\le(\sum_iR_i(z)^4)^{1/2}\le(KN)^{1/2}$. Because $N_1,N_0\ge cN$ for some $c>0$ and large $N$, $\max_i c_i^2=O(N^{-3/2})$ and $\max_i(c_i-\bar c)^2=O(N^{-3/2})$. By Assumption~\ref{ass:regularity}(c), $V_N\ge\kappa/N$ for some $\kappa>0$, so the ratio in Lemma~\ref{lem:rejective}(c) is $O(N^{-1/2})$, and part (a) follows.

\textsc{Step} 3 (variance estimator). Let $\bar R_1=N_1^{-1}\sum_iZ_iR_i$. Then $\bar R_1-m_1=N_1^{-1}\sum_iZ_ie_i(1)$ and
\[
\hat s^2_1=\frac{1}{N_1-1}\Bigl\{\sum_iZ_ie_i(1)^2-N_1(\bar R_1-m_1)^2\Bigr\}.
\]
First, $\ER[\sum_iZ_ie_i(1)^2]=\sum_i\pi_ie_i(1)^2$. Since $|m_1|\le N_1^{-1}\sum_i|R_i(1)|=O(1)$, $\sum_ie_i(1)^4=O(N)$, and Lemma~\ref{lem:rejective}(a) gives $\VR[N^{-1}\sum_iZ_ie_i(1)^2]\le(4N^2)^{-1}\sum_ie_i(1)^4=O(N^{-1})$. Second, $\ER[\sum_iZ_ie_i(1)]=\sum_i\pi_iR_i(1)-N_1m_1=0$ and, by Lemma~\ref{lem:rejective}(a), $\VR[\sum_iZ_ie_i(1)]\le\frac12\sum_ie_i(1)^2=O(N)$, so $\bar R_1-m_1=O_p(N^{-1/2})$. Therefore $\hat s_1^2=N_1^{-1}\sum_i\pi_ie_i(1)^2+O_p(N^{-1/2})$ and, analogously, $\hat s_0^2=N_0^{-1}\sum_i(1-\pi_i)e_i(0)^2+O_p(N^{-1/2})$. Since $\widehat V=\hat s_1^2/N_1+\hat s_0^2/N_0$ (expression \eqref{eq:Vhat} is the same quantity computed from the sample moments of $\Delta Y_i$ and $\Delta D_i$), $N(\widehat V-V_N^+)=O_p(N^{-1/2})$.

\textsc{Step} 4 (conservativeness). Let $k=m_1/N_1+m_0/N_0$, so that $c_i-k=e_i(1)/N_1+e_i(0)/N_0$. Direct expansion gives
\[
\frac{\pi_ie_i(1)^2}{N_1^2}+\frac{(1-\pi_i)e_i(0)^2}{N_0^2}-\tilde\pi_i\Bigl\{\frac{e_i(1)}{N_1}+\frac{e_i(0)}{N_0}\Bigr\}^2
=\Bigl\{\frac{\pi_ie_i(1)}{N_1}-\frac{(1-\pi_i)e_i(0)}{N_0}\Bigr\}^2.
\]
Summing over $i$, $V_N^+-\sum_i\tilde\pi_i(c_i-k)^2$ equals the right-hand side of \eqref{eq:conservative}. Since $\bar c$ minimizes $x\mapsto\sum_i\tilde\pi_i(c_i-x)^2$, $\sum_i\tilde\pi_i(c_i-k)^2\ge V_N$, which proves \eqref{eq:conservative}.

\textsc{Step} 5 (test validity). Let $G_N=(\widehat M-m_N)/\VR[\widehat M]^{1/2}$, so that $G_N\dto N(0,1)$ by Step 2. Under $m_N=0$, $T_N=G_N\{\VR[\widehat M]/\widehat V\}^{1/2}$. By Steps 2--4 and $NV_N\ge\kappa$,
\[
\frac{\widehat V}{\VR[\widehat M]}=\frac{V_N^++o_p(N^{-1})}{V_N\{1+o(1)\}}\ge\frac{V_N+o_p(N^{-1})}{V_N\{1+o(1)\}}=1+o_p(1).
\]
For any $\varepsilon\in(0,1)$, $\PR(|T_N|>z)\le\PR\{|G_N|>z(1-\varepsilon)\}+\PR\{\widehat V<(1-\varepsilon)^2\VR[\widehat M]\}$. The second term vanishes, so $\limsup_N\PR(|T_N|>z)\le2\{1-\Phi(z(1-\varepsilon))\}$. Letting $\varepsilon\downarrow0$ with $z=z_{1-\alpha/2}$ proves part (c).
\end{proof}

\begin{proof}[Proof of Corollary~\ref{cor:arset}]
Since $\tau_\FS\neq0$, $m_N(\beta_R)=0$, and Theorem~\ref{thm:inference}(c) at $b=\beta_R$ gives the coverage statement. Expanding $\widehat M(b)^2-\kappa\widehat V(b)$ with \eqref{eq:Vhat} gives $Ab^2-2Bb+C$. If $A>0$, the quadratic is convex and equals $-\kappa\widehat V(\widehat\beta)\le0$ at $b=\widehat\beta$, because $\widehat M(\widehat\beta)=0$; hence $B^2\ge AC$ and the set is the stated interval. If $A<0$, the quadratic is concave, and its sublevel set $\{\le0\}$ is the complement of the open interval between the real roots when $B^2>AC$ and the whole line when $B^2<AC$. Finally, $A>0$ if and only if $|\hatt_\FS|/\hat v_{DD}^{1/2}>z_{1-\alpha/2}$.
\end{proof}

\begin{proof}[Proof of Corollary~\ref{cor:wald}]
Since $\hatt_\RF-\beta_R\hatt_\FS=\widehat M(\beta_R)$, $\widehat\beta-\beta_R=\widehat M(\beta_R)/\hatt_\FS$, and the event $|\widehat\beta-\beta_R|>z\,\widehat{\mathrm{se}}$ is $|\widehat M(\beta_R)|>z\,\widehat V(\widehat\beta)^{1/2}$. Steps 2 and 3 of the proof of Theorem~\ref{thm:inference} applied to $\Delta D_i$ give $\hatt_\FS-\tau_\FS=O_p(N^{-1/2})$. In addition, $V_N(\beta_R)\le\frac14\sum_ic_i(\beta_R)^2=O(N^{-1})$, so $\widehat M(\beta_R)=O_p(N^{-1/2})$ and $\widehat\beta-\beta_R=O_p(N^{-1/2})$. Since $\hat v_{YY},\hat v_{DD}=O_p(N^{-1})$ and $|\hat v_{YD}|\le(\hat v_{YY}\hat v_{DD})^{1/2}$, $\widehat V(\widehat\beta)-\widehat V(\beta_R)=-2(\widehat\beta-\beta_R)\hat v_{YD}+(\widehat\beta^2-\beta_R^2)\hat v_{DD}=o_p(N^{-1})$. Step 5 of the proof of Theorem~\ref{thm:inference} then applies with $\widehat V(\widehat\beta)$ in place of $\widehat V(\beta_R)$.
\end{proof}

\subsection{Proofs for Section~\ref{sec:sensitivity}}

\begin{proof}[Proof of Proposition~\ref{prop:idset}]
By \eqref{eq:betaC-identity}, $\beta_C=g(R_N,F_N)$ with $g(r,f)=(\tau_\RF-r)/(\tau_\FS-f)$, $|R_N|\le M_R$, $|F_N|\le M_F$ and $\tau_\FS-F_N=B_\cC>0$; hence $\beta_C\in\Theta_I$. If $M_F<\tau_\FS$, $g$ is continuous on the compact and connected box $[-M_R,M_R]\times[-M_F,M_F]$, so its image is a compact interval. For fixed $f$, $g$ is strictly decreasing in $r$. For fixed $r$, $\partial g/\partial f=(\tau_\RF-r)/(\tau_\FS-f)^2$ has constant sign, so $g$ is monotone in $f$. The minimum is therefore attained at $r=M_R$ and $f\in\{-M_F,M_F\}$, and the maximum at $r=-M_R$ and $f\in\{-M_F,M_F\}$.
\end{proof}

\begin{proof}[Proof of Theorem~\ref{thm:sens}]
By \eqref{eq:betaC-identity}, $m_N(\beta_C)=\tau_\RF-\beta_C\tau_\FS=R_N-\beta_CF_N$, so $|m_N(\beta_C)|\le M_R+|\beta_C|M_F$ under Assumption~\ref{ass:bounds}. Hence
\[
\PR\bigl\{\beta_C\notin\mathrm{CS}^S_\alpha(M_R,M_F)\bigr\}\le\PR\Bigl\{|\widehat M(\beta_C)-m_N(\beta_C)|>z_{1-\alpha/2}\widehat V(\beta_C)^{1/2}\Bigr\}.
\]
Steps 1--5 of the proof of Theorem~\ref{thm:inference}, applied at $b=\beta_C$ to the centred statistic $\widehat M(\beta_C)-m_N(\beta_C)$, show that the right-hand side has limit superior at most $\alpha$. The case $M_R=M_F=0$ is immediate from the definitions.
\end{proof}

\begin{proof}[Proof of Corollary~\ref{cor:breakdown}]
At $b=0$, $\widehat M(0)=\hatt_\RF$, $\widehat V(0)=\hat v_{YY}$ and $|b|M_F=0$.
\end{proof}

\begin{proof}[Proof of Proposition~\ref{prop:known}]
(a) Under Assumption~\ref{ass:strata}, $\ER[\bar X_{zg}]=N_g^{-1}\sum_{i\in\mathcal G_g}X_i(z)$. Hence
\[
\ER[\hat\theta_X]=\sum_g\sum_{i\in\mathcal G_g}\frac{N_{1g}}{N_1N_g}\{X_i(1)-X_i(0)\}=\sum_ib_i\{X_i(1)-X_i(0)\},
\]
since $\pi_i=N_{1g}/N_g$ for $i\in\mathcal G_g$, and similarly
\[
\ER[\hat\rho_X]=\sum_g\sum_{i\in\mathcal G_g}\Bigl(\frac{\pi_g}{N_1}-\frac{1-\pi_g}{N_0}\Bigr)X_i(0)=\sum_ia_iX_i(0).
\]
By Table~\ref{tab:types}, $\sum_ib_i\{\Delta D_i(1)-\Delta D_i(0)\}=B_\cC$, $\sum_ib_i\{\Delta Y_i(1)-\Delta Y_i(0)\}=B_\cC\beta_C$, $\sum_ia_i\Delta D_i(0)=F_N$ and $\sum_ia_i\Delta Y_i(0)=R_N$. The identity $\hatt_X=\hat\theta_X+\hat\rho_X$ follows from $\hatt_X=\sum_g\{(N_{1g}/N_1)\bar X_{1g}-(N_{0g}/N_0)\bar X_{0g}\}$, and the expression for $\beta_R-\beta_C$ is Theorem~\ref{thm:decomp}(b).

(b) Under $H_0:\beta_C=b$, part (a) gives $\ER[\widetilde M(b)]=B_\cC(\beta_C-b)=0$. Write $\widetilde M(b)=\sum_gw_g(\bar R_{1g}-\bar R_{0g})$ with $w_g=N_{1g}/N_1$ and $\bar R_{zg}$ the mean of $R_i(b)$ in stratum $g$ and arm $z$. The summands are independent across strata. Within each stratum, assignment is completely randomized with $N_{1g}/N_g\in[\eta,1-\eta]$, and the moment bound implies the conditions of the finite-population central limit theorem of \citet[][Theorem 1]{LiDing2017}. Since $G$ is fixed and $N\VR[\widetilde M(b)]$ is bounded away from zero, $\widetilde M(b)/\VR[\widetilde M(b)]^{1/2}\dto N(0,1)$. Within stratum $g$, $\VR[\bar R_{1g}-\bar R_{0g}]=S^2_{1g}/N_{1g}+S^2_{0g}/N_{0g}-S^2_{\tau g}/N_g$, where $S^2_{zg}$ is the finite-population variance of $R_i(z;b)$ in stratum $g$ and $S^2_{\tau g}$ that of $R_i(1;b)-R_i(0;b)$; and $\hat s^2_{zg}(b)-S^2_{zg}\pto0$ \citep{LiDing2017}. Hence $N\{\widetilde V(b)-\VR[\widetilde M(b)]\}\ge o_p(1)$, and the argument of Step 5 of the proof of Theorem~\ref{thm:inference} completes the proof.
\end{proof}
}{%
  \PackageWarningNoLine{main}{appendix_proofs.tex not found; appendix omitted}%
}

\end{document}